\documentclass[aps,prd,twocolumn,superscriptaddress,showpacs]{revtex4}
\usepackage{graphicx}
\usepackage{dcolumn}
\usepackage{bm}
\usepackage{natbib}

\newcommand{\be}{\begin{equation}}
\newcommand{\ee}{\end{equation}}
\newcommand{\bea}{\begin{eqnarray}}
\newcommand{\eea}{\end{eqnarray}}

\begin{document}
%opening
\title{Chaplygin gas in light of recent integrated~Sachs--Wolfe~effect data}

\author {Tommaso Giannantonio}
\email {tommaso.giannantonio@port.ac.uk}

\affiliation {Institute of Cosmology and Gravitation, University
of Portsmouth, Mercantile House, Hampshire Terrace, PO1 2EG, Portsmouth, England.}

\author {Alessandro Melchiorri}
\email {alessandro.melchiorri@roma1.infn.it}

\affiliation {Dipartimento di Fisica ``G. Marconi'', Universit\`a
di Roma ``La Sapienza'', P.le Aldo Moro 5, 00185, Rome, Italy.}
\affiliation {INFN, sezione
  di Roma, Universit\`a di Roma ``La Sapienza'', P.le Aldo Moro 5,
  00185, Rome, Italy.}

\begin {abstract}
We investigate the possibility of constraining Chaplygin dark energy 
models with current Integrated Sachs Wolfe effect data.
In the case of a flat universe we found that 
generalized Chaplygin gas models must have an energy density 
such that $\Omega_c >0.55$ and an equation of state
$w <-0.6$ at $95 \%$ c.l.. We also investigate the recently proposed
Silent Chaplygin models, constraining $\Omega_c >0.55$ and 
$w <-0.65$ at $95 \%$ c.l..
Better measurements of the CMB-LSS correlation will be possible
with the next generation of deep redshift surveys. This will provide
independent and complementary constraints on unified dark energy models such as the
Chaplygin gas.

\end {abstract}
\pacs {}
\maketitle

\section {Introduction} \label {sec:intro}

A decade of impressive progress in cosmology has finally led to the
so-called $ \Lambda $CDM model of structure formation,
based on inflation, cold dark matter and a cosmological constant
(see e.g. \cite{spergel}). However, this can't be seen as the 
ultimate result, since its features include a massive 
amount of energy of unknown origin in the universe, usually split in 
two categories called dark matter (DM) and dark energy (DE)
(see e.g. \cite{cosmo}).

In fact, there are strong experimental constraints that have induced 
the introduction of new items in the cosmic energy balance, and these 
constraints arise from two different classes of observations: the ones 
based on the study of the property of the matter and its clustering 
have suggested the presence of DM, while the study 
of the cosmic microwave background (CMB) and supernovae Ia (SNaeIa) 
has led to a wide consensus about DE.

Nevertheless, our theoretical knowdledge of the dark universe is still 
quite poor since it is not directly detected but only 
inferred from its cosmological implications, and many models for 
DM and DE are not yet ruled out by the observations. 
For the sake of simplicity, one could then make the ansatz that 
these two entities are in reality just two distinct manifestations of 
one thing. This is the purpose for which have been introduced the 
unified dark energy models (UDE), also called quartessence.

An alternative to quintessence model would be an exotic fluid capable 
to develop a negative pressure at late times, while approximating the 
dust matter behaviour earlier on, with a smooth transition: these are 
exactly the features of the Chaplygin gas, a fluid introduced in 1904 
\cite {Chaplygin:1904} for aerodynamics, that presents some 
interesting properties recently discovered for cosmology 
\cite{Kamenshchik:2001cp}.

The Chaplygin gas is characterized by an equation of state of the form
\be \label {eq:statechap}
p_X = - \frac {A} {\rho_X^{\alpha}},
\ee
with $ \alpha = 1 $, while a generalized Chaplygin gas has $ 0 <
\alpha \le 1 $ \cite {Bento:2002ps}, 
and $ A $ is a constant with dimensions $ M^{4 (1 + \alpha)} $. 

The Jeans lenght of the GCG perturbations is first similar to the
matter one, and then the instability is removed when the behaviour 
approximates a cosmological constant; this produces a large integrated 
Sachs--Wolfe (ISW) effect \cite{Carturan:2002si} on the CMB anisotropies, 
significatively different from the one expected in $\Lambda$CDM models.
Recent combined analysis of CMB anisotropies and Large Scale structure data
have severly constrained the Chaplygin gas model (see for example 
\cite{Bean:2003fb}). 
However, several cosmological backgrounds may mimick an enhancement in the
large angular scale anisotropy as the one produced by the ISW 
like, just to name a few, gravitational waves or cosmic strings.
It is therefore important to verify this result by using independent and
complementary datasets.
 
A powerful and model-independent way to extract the ISW signal from large scale
CMB anisotropy has been proposed in \cite{Crittenden:1995ak} by cross-correlating
CMB maps with galaxy surveys. With the advent of the new WMAP results, several
cross-correlation analysis have been made with at least five detections
at $>2 \sigma$ level. It is therefore timely to investigate the impact of
those measurements on the Chaplygin gas model.
In the following of the paper, we use the recent detections of 
the ISW effect in the CMB anisotropies resumed in \cite{Gaztanaga:2004sk}.

In the next section we will discuss the Chaplygin gas model and
the expected ISW signal. In section 3 we will analyse the current data
and produce new independent constraints on the model and finally in section
4 we will derive our conclusions.

\section {The Chaplygin gas and the integrated Sachs--Wolfe effect}
\label {sec:isw-cha}

From equation \ref{eq:statechap} and energy
conservation it follows that for the Chaplygin 
(see e.g. \cite {Amendola:2003bz})

\be
\rho_X = \left[ A + \frac {B} {a^{3 (\alpha + 1)}} \right]^{\frac {1} {1 + \alpha}},
\ee
where $ B $ is a constant with the same dimensions of $ A $; this
means that the equation of state for the GCG is of the form
\be
w (a) \equiv \frac {p_X} {\rho_X} = - \frac {A} {\rho_X^{1 + \alpha}},
\ee
that at present time is
\be
w = - \frac {A} {A + B}.
\ee

\noindent Using the physical parameters $ \Omega_X $ and $ w $ instead of $ A
$ and $B$, we can recover the expression of the evolving equation of state:

\be
w (a) = \left[ \frac {1 + w} {w a^{3 (\alpha + 1)}} - 1 \right]^{-1},
\ee

\noindent From the equation above it is clear that at early times
the fluid behaves as non relativistic matter while at late
times it behaves as a fluid with equation of state
$w$. 

\noindent The corresponding density evolution equation is
\be
\rho (a) = \left[ -w + \frac {1 + w} {a^{3 (\alpha + a)}} \right]^{\frac {1} {1 + \alpha}}.
\ee

In the synchronous gauge the evolution equations for 
the density and velocity divergence perturbations, $\delta$ and
$\theta$, in Fourier space for the Chaplygin gas  are (see e.g. \cite{beanchap})

\be \label{eq:delta}
\dot \delta = - (1 + w) \left( \vartheta + \frac {\dot h} {2} \right)
+ 3 {\cal H} w (1+\alpha) \delta
\ee

\be \label {eq:theta}
\dot \theta = -(1+3w \alpha){\cal H}\theta- { {\alpha w} \over {1+w}} k^2 \delta
\ee

\noindent where the derivatives are respect to the conformal time $(d/d \tau)$ and
$a {\cal H}=da/d \tau$. As showed in \cite{Bean:2003fb} below a characteristic 
scale $k_*^2 > {\cal H}^2/(|\alpha w|)$ $\delta$ has oscillating (growing) solutions
for $\alpha > 0$ ($\alpha < 0$). This behaviour makes the Chaplygin gas
at odds with the current cosmological data when considered as unified dark matter
model. The generalized Chaplygin gas can therefore be considered only as
a candidate for the Dark Energy component.
Recently, Amendola et al. \cite{Amendola:2005rk} have introduced a new 
version of the Chaplygin
model, whith an additional entropy perturbation component $\Gamma$ 
such that:

\be
{{\delta P} \over {\delta \rho}}= {p \over {\delta \rho}} \Gamma + c^2_s=0
\ee

\noindent i.e. the effective sound speed of the cosmic fluid vanishes,
assuring clustering on small scales. 
In this case, the perturbation equations change to

\be
\dot \delta = - (1 + w) \left( \vartheta + \frac {\dot h} {2} \right)
+ 3 {\cal H} w \delta
\ee

\be
\dot \theta = -(1+3w \alpha){\cal H}\theta
\ee

The Silent Chaplygin model is in good agreement with the
current status of observations \cite{Amendola:2005rk}.

Differences between Chaplygin models
may arise at late times, when their energy contribution becomes 
important; for this reason,  it is useful to study the 
consequences of both models 
on a typically late time phenomenon as the 
integrated Sachs--Wolfe effect \cite{Sachs:1967er}.

When CMB photons pass through potential wells they can acquire 
a red-- or blue--shift if the gravitational
potential is not constant in time; 
this is exactly what happens when in the universe energy density
balance dark energy 
becomes to be important. This effect can only arise at late times,
when $ z < 4 $ (and $\Omega_m$ significantly different from one), and when 
galactic structures have already been formed; for this reason, their
distribution 
will be correlated with the ISW signal, while the primary CMB
anisotropy signal can't. 
It's for this reason that the best way to measure the ISW signal in
the CMB anisotropies spectrum is the cross--correlation technique between the
whole signal and a survey of the matter distribution in the universe 
\cite {Crittenden:1995ak}. The ISW temperature fluctuation in the direction
 $ \hat {\textbf {n}} $ is given by
\be
\Theta_{ISW} (\hat {\textbf {n}}) = -2 \int e^{-\tau (z)} \frac {d \Phi} {dz} (\hat {\textbf {n}}, z) dz,
\ee
where $ \Phi $ is the newtonian gauge gravitational potential and $ e^{-\tau (z)} $
is the visibility function to account for a possible suppression due to early reionization.
Conversely, the observed overdensity in a given direction is
\be
\delta_{LSS} (\hat {\textbf {m}}) = b_g \int \varphi (z) \delta_m (\hat {\textbf {m}}, z) dz,
\ee
where $ \delta_m $ is the matter density perturbation, $ b_g $ the
galactic bias and $ \varphi (z) $ the specific survey selection
function. With these two definitions, one can compute the 2--points angular cross-correlation function
\be
c^{ISW-LSS} (\vartheta) = \langle \Theta_{ISW} (\hat {\textbf {n}}) \delta_{LSS} (\hat {\textbf {m}}) \rangle,
\ee
where $ \vartheta $ is the angle between the directions $ \hat {\textbf {n}} $ and  $ \hat {\textbf {m}} $, and the cross--correlation power spectrum \cite {Garriga:2003nm}
\be
C_l^{ISW-LSS} = 4 \pi \frac {9} {25} \int \frac {dk} {k} \Delta_R^2 I_l^{ISW} (k) I_l^{LSS} (k),
\ee
where $ \Delta_R^2 $ is the primordial power spectrum and
\bea
I_l^{ISW} (k) & = & -2 \int e^{-\tau (z)} \frac {d \Phi_k} {dz} j_l [kr (z)] dz \\
I_l^{LSS} (k) & = & b_g \int \varphi (z) \delta^k_m (z) j_l [kr (z)] dz,
\eea
where $ \Phi_k $ and $ \delta^k_m $ are the Fourier components of the gravitational potential and matter perturbation, $ r (z) $ is the comoving distance at redshift $ z $ and  $ j_l [kr (z)] $ are the spherical Bessel functions.

In Figure 1 we plot some theoretical $ISW-LSS$ power spectra
computed under the case of generalized and silent chaplygin gases
respectively. We consider the Chaplyging gas as
a candidate for dark energy alone, we fix $\Omega_m=0.3$, $w=-0.9$ 
and we let vary $\alpha$. The prediction is made for an all-sky CMB
survey plus a LSS survey centered at redshift $z_*=0.3$ with dispersion
$\Delta z \sim 0.1$. 
In the top panel we consider predictions for the Generalized
Chaplygin. As we can see there is a strong, non trivial, dependence
on $\alpha$. Increasing $\alpha$ shiftss the epoch of matter to
dark energy transition, affecting the amplitude of the ISW signal.
At the same time, the Chaplygin gas affects the growth of
perturbations in the CDM component, decreasing the growth
in the case of $\alpha=0$. In the $ISW-LSS$ cross signal the
two mechanism are in competition and the signal is strongly dependent
from the value of $\alpha$ and from the details ofthe LSS
survey. On the bottom panel we plot the same predictions but in the
case of the silent Chaplygin. As we can see, the additional intrinsic
entropy perturbation has the effect of making the cross signal
substantially independent from $\alpha$. 

 \begin{figure}[t]
\begin{center}
\includegraphics[angle=-90,width=1.05\linewidth]{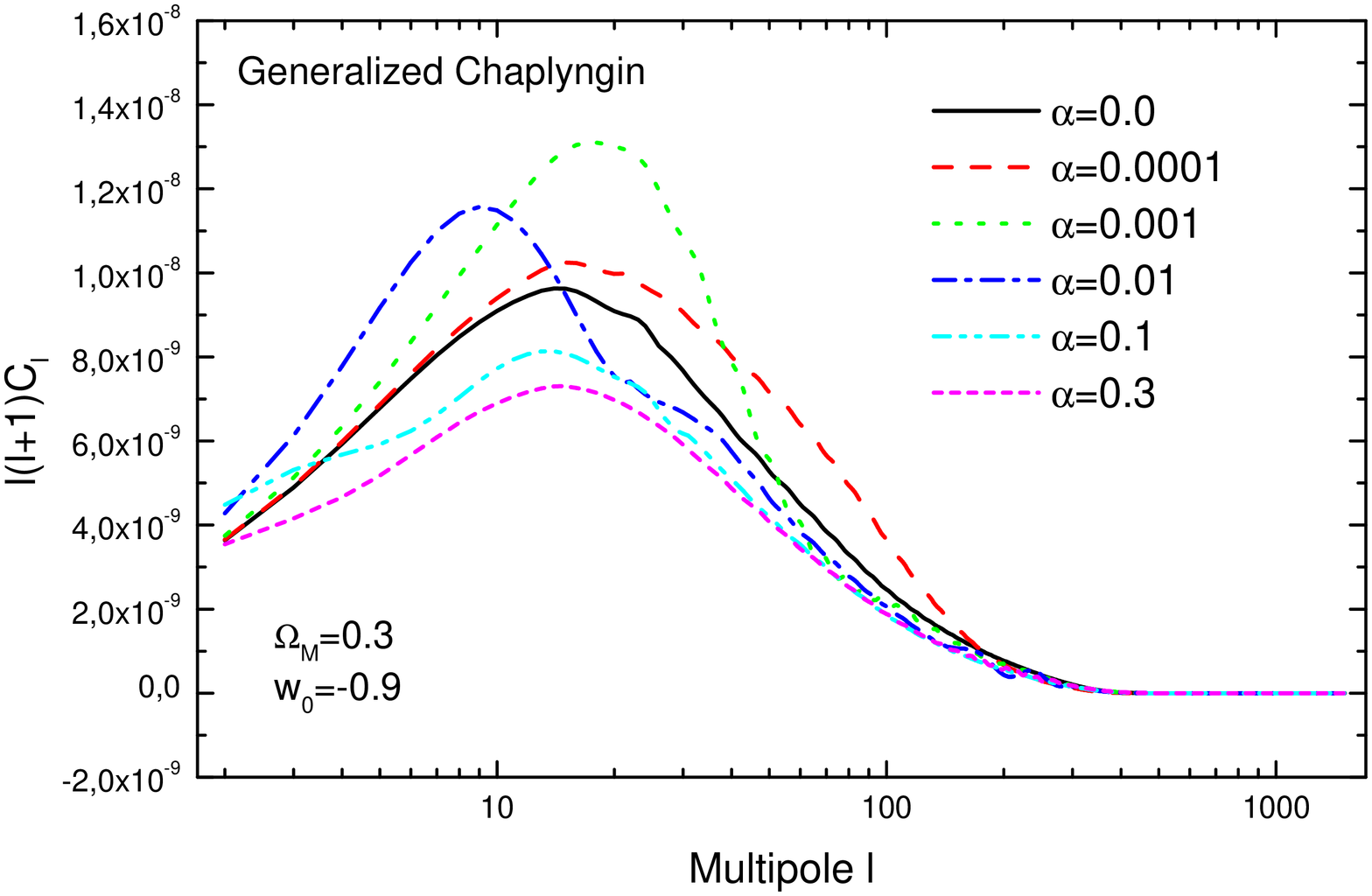}
\includegraphics[angle=-90,width=1.05\linewidth]{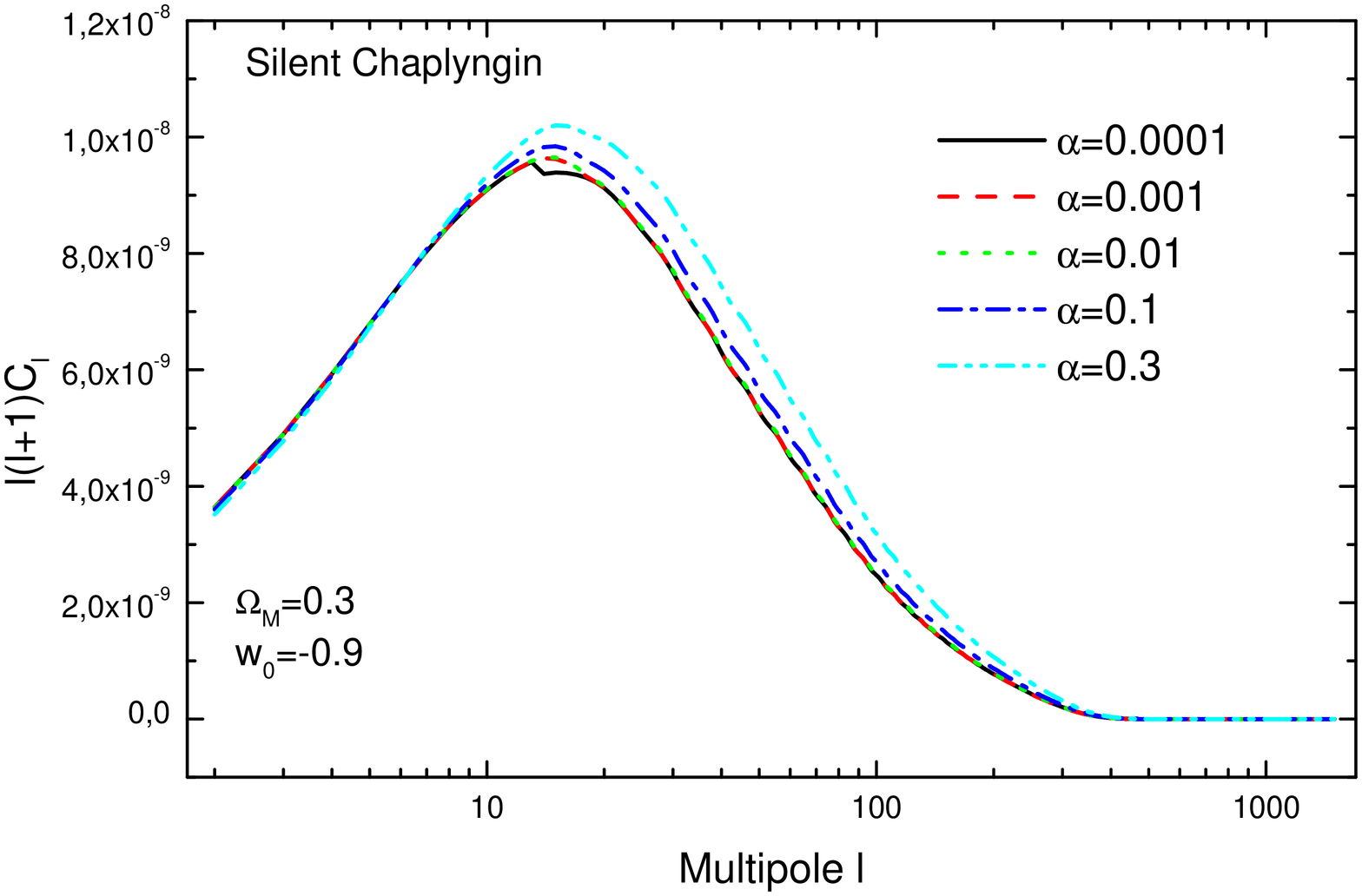}

\caption{Predictions for the $ISW-LSS$ power spectrum in the
case of the generalized (top panel) and silent (bottom panel)
Chaplygin. 
We fix $\Omega_m=0.3$ and $w=-0.9$. The LSS survey is centered
at $z_*=0.3\pm0.1$. }
\end{center}
\end{figure}

\section {Constraints from data}
\label {sec:constr}

We perform a likelihood analysis using the collection 
of data presented in Gaztanaga {\em et al.} \cite{Gaztanaga:2004sk}, 
which has the advantage of being publicly available and easy to implement.
In order to avoid degeneracies between the parameters we consider 
only two possible viable Chaplygin dark energy models:
a generalized Chaplygin model with $\alpha=0$ and the silent
quintessence model. For those models we vary two parameters:
the energy density $\Omega_{Chap}$ and the equation of state
$w$.

The data we consider consist of measurements of the average 
angular cross-correlation around $\theta=5^{\circ}$ between 
WMAP temperature anisotropy maps and several LSS surveys. 
The angular average around $\theta=5^{\circ}$
ensures that the signal is not 
contaminated by foregrounds such as the
SZ or lensing effects which are relevant at smaller angles 
($\theta\lesssim 1^{\circ}$).  
Possible systematic contaminants, 
such as extinction effects, seem not to affect these data and 
for a more detailed discussion we refer to \cite{Gaztanaga:2004sk}.
The data span a range of redshift $0.1<z\lesssim 1$ and
for each redshift bin the data include an estimate of the galaxy bias
with $20\%$ errors. These biases are inferred by comparing the 
galaxy-galaxy correlation function of each experiments 
with the expectation of best fit model to WMAP power spectra.
There is little sensitivity to the scalar spectral index $n_s$, while
the dependence on the baryon density $\Omega_b$ can be non-negligible. 
In fact the presence of baryons inhibites the growth of CDM fluctuations
between matter-radiation equality and photon-baryon decoupling
causing the matter power spectrum to be suppressed on scales $k>k_{eq}$
for increasing values of $\Omega_b$
($k_{eq}$ is the scale which enters the horizon at the equality). 
Over the range of scales which
contribute to the ISW-correlation ($k\sim 0.01$) 
the sensitivity on $\Omega_b$ is still present.
In order to limit the number of likelihood parameters
we therefore assume a Gaussian prior on the baryon density 
$\Omega_b =0.04$ and we take the Hubble parameter as
$h=0.7$. We also assume a scale invariant primordial spectrum $n_s=1$ and fix
the optical depth to reionization to WMAP best fit value
$\tau=0.17$ (again the ISW is not particularly affected by a change
in those parameters). We marginalize over the normalization amplitude
$A_s$, although we found no difference assuming the WMAP value. 
In fact changing $A_s$ shifts the overall amplitude of the angular 
cross-correlation of the same amount over different redshifts but it 
does not change the redshift dependence of the signal.
Since the experimental data are corrected for the bias 
by comparing the measured galaxy-galaxy correlation function
in each redshift bin to the WMAP best fit model, 
we rescale these biases to each of the dark energy model
in our database as described in \cite{Gaztanaga:2004sk}
and \cite{cora}.

We compute for each theoretical model
the angular cross-correlation as described in the previous section
using the selection function 
\begin{equation}
\phi^i(z)\sim z^2 \exp{[-(z/\bar{z}_i)^{1.5}]},\label{sel}
\end{equation} 
where $\bar{z}_i$ is the median redshift of the $i$-th survey.
Then following \cite{Gaztanaga:2004sk}, we compute the average cross-correlation
in the $i$-th bin, $\bar{C}_i^X$,
around $\theta\sim 5^{\circ}$.
 
The data points are an average over
angles and are infered from surveys whose selection functions
may overlap in redshift space, hence they are not independent
measurements and indeed are affected by a certain degree of correlation.
Since we have no access to the raw data 
we have no way of accounting for the first type of correlation, while
using Eq.~(\ref{sel}) we can estimate the correlation between
different redshift bins. We compute the correlation matrix 
$\rho=\{\rho_{ij}\}$, where $\rho_{ij}$ is the fraction 
of overlapping volume between the $i$-th and $j$-th surveys (i.e. the diagonal
components are $\rho_{ii}=1$). 
To be more conservative we have assumed that the different surveys 
cover the same fraction of sky, in general this is not the case
and the fraction of overlaping volume can be smaller.
We found that only two data points are highly correlated,
since their selection functions overlap for about $70\%$ in redshift space, 
while the correlation among the remaining data points are less than $20\%$.

We compute a likelihood function $\cal L$ 
defined as

\begin{equation}
-2\log{\cal L}=\chi^2=\sum_{ij}(\bar{C}_i^X-{\hat C}_i^X)
M_{ij}^{-1}(\bar{C}_j^X-{\hat C}_j^X),
\end{equation}
where ${\hat C}^X_i$ are the data and
$M_{ij}^{-1}=\rho_{ij}/(\sigma_i \sigma_j)$ is our estimate of 
the inverse of the covariance matrix, which takes in to account the
experimental errors and possible correlation amongst the data,
with $\sigma_i$ the measured 
uncertainty in the $i$-th bin. 

 \begin{figure}[t]
\begin{center} 
\includegraphics[width=0.95\linewidth]{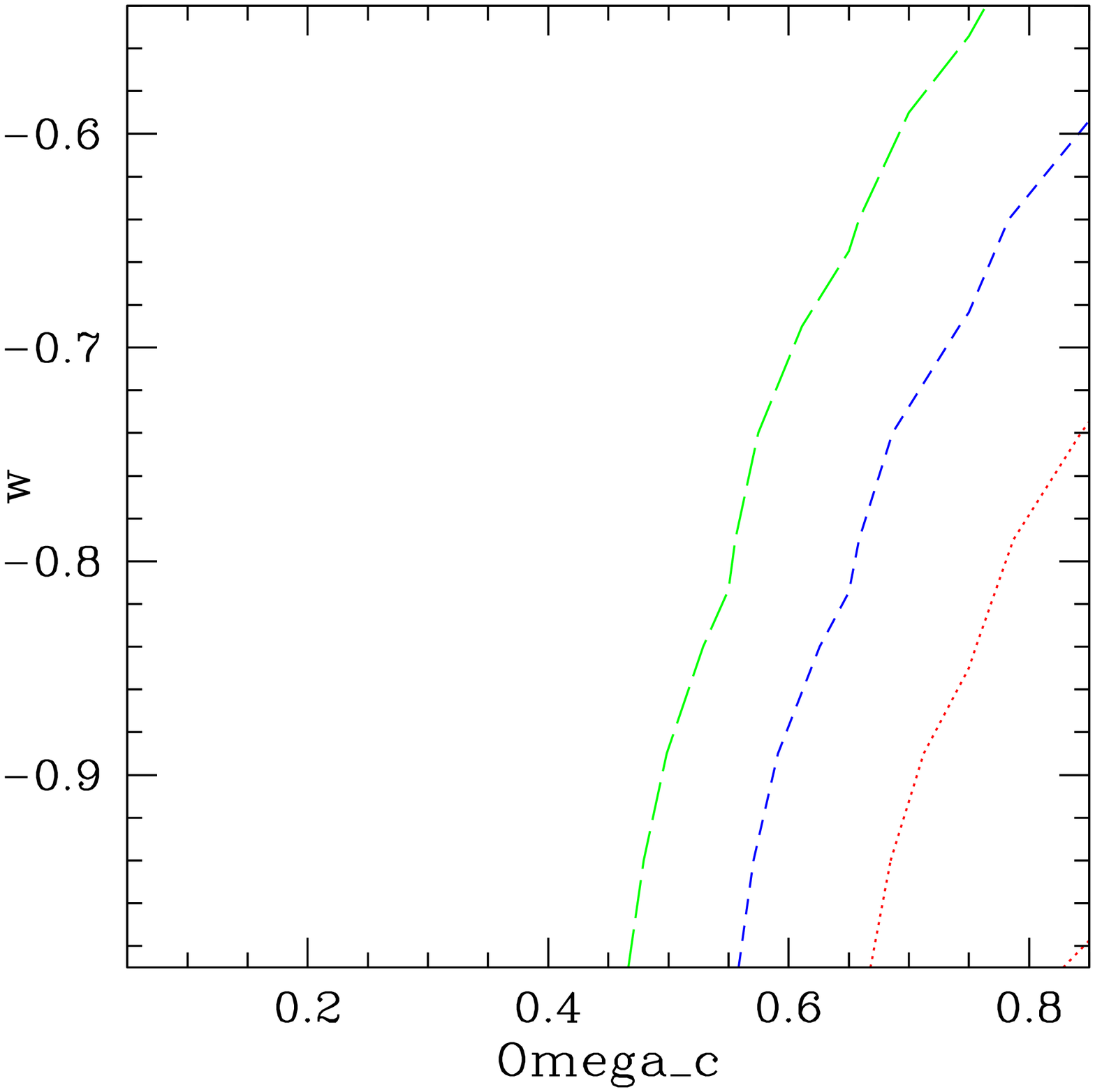}
\includegraphics[width=0.95\linewidth]{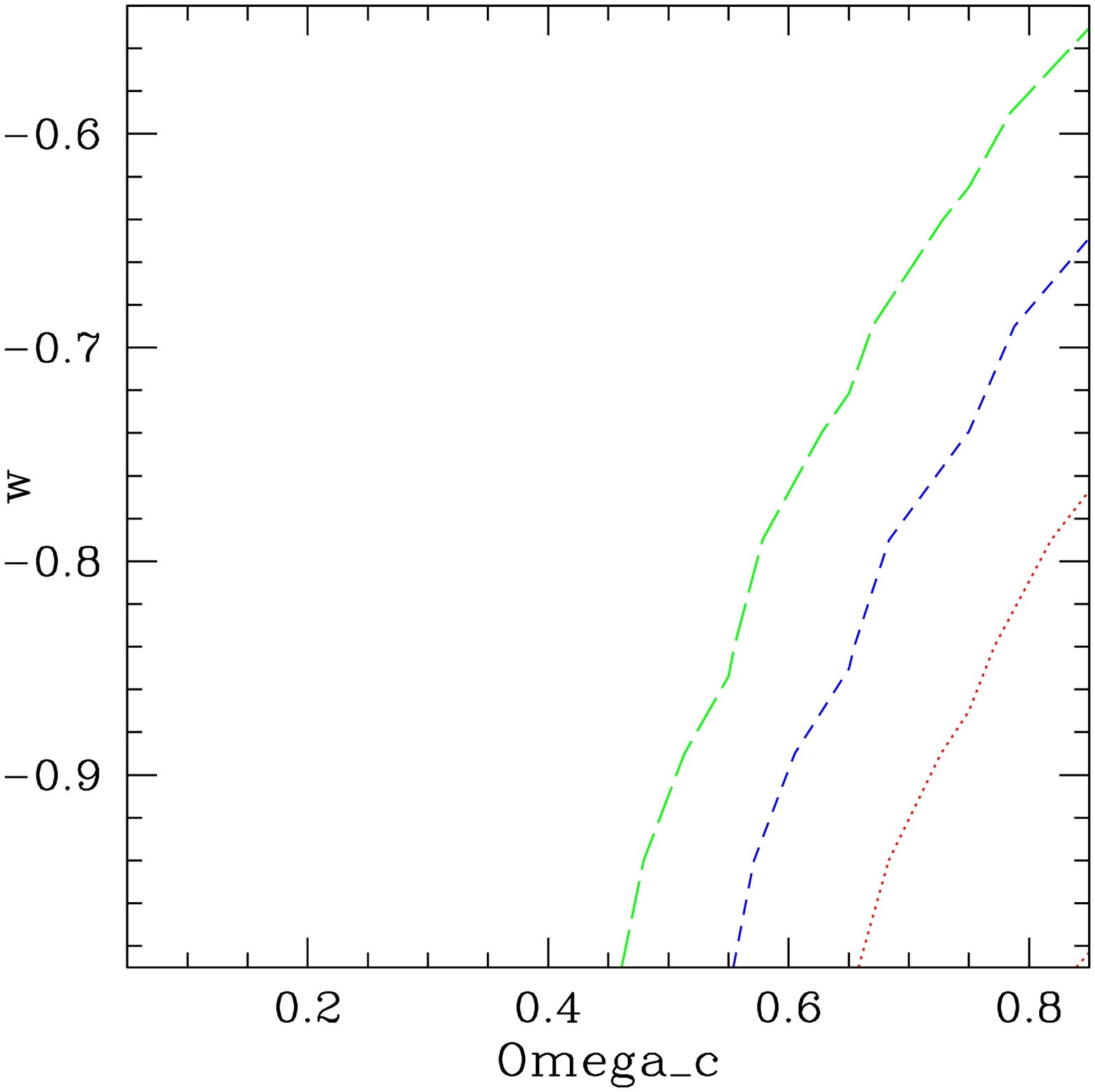}

\caption{Likelihood contour plots in the $\Omega_c$-$w$ plane for
a generalized Chaplygin gas with $\alpha=0$ (Top Panel) and
for the silent Chaplygin case (Bottom Panel).
The contours are $68 \%$ (dotted), $95 \%$ (short-dashed),
$99 \%$ (long-dashed) c.l..}
\end{center}
\end{figure}

The results of our analysis are plotted in Fig. 2 As we can see,
in the case of the generalized Chaplygin with $\alpha=0$ the current 
data constrain $\Omega_c > 0.55$ and $w <-0.6$ at $95 \%$ c.l..
Considering the silent Chaplygin model we obtain similar results
with $\Omega_c > 0.55$ and $w < -0.65$ again at $95 \%$ c.l..

\section {Conclusions}
\label {sec:conc}

In this paper we have investigated the possibility of constraining 
Chaplygin models with the current Integrated Sachs Wolfe effect data.
In the case of a flat universe we found that cosmological viable 
generalized Chaplygin gas models must have an energy density 
such that $\Omega_c >0.55$ and an equation of state
$w <-0.6$ at $95 \%$ c.l.. We also investigate the recently proposed
Silent Chaplygin models, constraining $\Omega_c >0.55$ and 
$w <-0.65$ at $95 \%$ c.l..
These constraints have been obtained with a prior on the
value of the baryon density $\Omega_b$ in agreement with BBN 
nucleosynthesis. 
Better measurements of the CMB-LSS correlation will be possible
with the next generation of deep redshift surveys. 
For an ideal whole sky experiment the ISW signal may be detected at
the level of $\sim 10 \sigma$. This would translate in a 
factor $2$/$3$ better determination of the results presented here
and will provide
independent and complementary constraints on unified dark 
energy models such as the Chaplygin gas.

\section{Acknowledgments}

We thank Pier Stefano Corasaniti
and Rachel Bean for valuable help and discussions.

\newpage
~
\newpage

\end {document}